\documentclass[draft,apl,showpacs,floatfix,12pt]{revtex4}
\usepackage{amssymb}
\usepackage{amsmath}
\usepackage{epsfig}
\usepackage{graphics}

\begin{document}

\title{A Datta-Das transistor with enhanced spin control}
\author{J. Carlos \surname{Egues$^{a)}$}}
\author{Guido Burkard}
\author{Daniel Loss}
\affiliation{Department of Physics and Astronomy, University of
Basel, Klingelbergstrasse 82, CH-4056 Basel, Switzerland}

\begin{abstract}
We consider a two-channel spin transistor with weak spin-orbit induced
interband coupling. We show that the coherent transfer of carriers between
the coupled channels gives rise to an \textit{additional} spin rotation. We
calculate the corresponding spin-resolved current in a Datta-Das geometry
and show that a weak interband mixing leads to enhanced spin control.
\end{abstract}

\date{\today }
\pacs{71.70.Ej,72.25.-b,73.23.-b,73.63.Nm}
\maketitle

\vfill $^{a)}$\small{Also at: Department of Physics and
Informatics, University of S\~ao Paulo at S\~ao Carlos, 13560-970
S\~ao Carlos/SP, Brazil; electronic mail: egues@if.sc.usp.br}

\newpage

The pioneering spin-transistor proposal of Datta and Das \cite{datta} best
exemplifies the relevance of electrical control of magnetic degrees of
freedom as a means of spin modulating charge flow. In this device \cite{exp}%
, a spin-polarized current \cite{spin-pol,egues} injected from the source is
spin modulated on its way to the drain via the Rashba spin-orbit \cite%
{rashba} (s-o) interaction, Fig. 1(a). The spin transistor operation relies
on gate controlling \cite{nitta} the strength $\alpha $ of the Rashba
interaction which has the form $H_{R}=i\alpha \sigma _{y}\partial /\partial
x $ in a strictly 1D channel \cite{rashba}. Upon crossing the Rashba-active
region of length $L$, a spin-up incoming electron emerges in the
spin-rotated state%
\begin{equation}
\left(
\begin{array}{c}
1 \\
0%
\end{array}%
\right) \rightarrow \left(
\begin{array}{c}
\cos (\theta _{R}/2) \\
-\sin (\theta _{R}/2)%
\end{array}%
\right) ,  \label{eq1}
\end{equation}%
where $\theta _{R}=2m^{\ast }\alpha L/\hbar ^{2}\equiv 2k_{R}L$ is
the rotation angle and $m^*$ is the electron effective mass
\cite{datta}. The corresponding spin-resolved conductance is found
to be $G_{\uparrow ,\downarrow }=e^{2}(1\pm \cos \theta _{R})/h$.

Here we extend the above picture by considering a geometry with two
weakly-coupled Rashba bands in the quasi-one-dimensional channel, Fig. 1(b).
We treat the degenerate $k$ states near the band crossings perturbatively in
analogy to the nearly-free electron model \cite{am}. This approach allows
for a simple analytical description of the problem. We calculate the
spin-resolved current by extending the usual procedure of Datta and Das \cite%
{datta} to account for weakly coupled bands. Our main finding is an \emph{%
additional} spin rotation for injected electrons with energies near the band
crossing [see shaded region around $\varepsilon _{F}$ in Fig. 2]. As we
derive later on, an incoming spin up electron in channel \emph{a} emerges
from the Rashba region in the rotated state%
\begin{equation}
\left(
\begin{array}{c}
1 \\
0 \\
0 \\
0%
\end{array}%
\right) \rightarrow \frac{1}{2}\left(
\begin{array}{c}
\cos (\theta _{d}/2)e^{-ik_{R}L}+e^{ik_{R}L} \\
-i\cos (\theta _{d}/2)e^{-ik_{R}L}+ie^{ik_{R}L} \\
-i\sin (\theta _{d}/2)e^{-ik_{R}L} \\
\sin (\theta _{d}/2)e^{-ik_{R}L}%
\end{array}%
\right) ,  \label{eq2}
\end{equation}%
where $\theta _{d}=\theta _{R}d/k_{c}$ is the additional spin rotation
angle, $d$ the interband matrix element and $k_{c}$ the wave vector at the
band crossing, Fig. 2. From (\ref{eq2}) we can find the new spin-resolved
conductance
\begin{equation}
G_{\uparrow ,\downarrow }=\frac{e^{2}}{h}\left(
\begin{array}{c}
1+\cos \left( \theta _{d}/2\right) \cos \theta _{R} \\
1-\cos \left( \theta _{d}/2\right) \cos \theta _{R}%
\end{array}%
\right) .  \label{eq3}
\end{equation}%
We now proceed to derive Eqs (\ref{eq2}) and (\ref{eq3}).

\emph{Model.} We consider a quasi-one-dimensional wire of length $L$ with
two bands \emph{a} and \emph{b}\textbf{\ }described by $\varepsilon
_{n,\sigma _{z}}(k)=\hbar ^{2}k^{2}/2m^{\ast }+\epsilon _{n},$ \ $n=a,b$ and
eigenfunctions $\varphi _{k,n,\sigma }(x,y)=e^{ikx}\phi _{n}(y)|\sigma
\rangle /\sqrt{L},$ $\sigma =\uparrow ,\downarrow $ where the $\phi _{n}(y)$%
's denote the transverse confinement wave functions. In the presence of the
Rashba s-o interaction, we can derive a Hamiltonian for the system in the
basis of the uncoupled wave functions $\{\varphi _{k,n,\sigma _{z}}(x,y)\}$.
This reads, 
\begin{equation}
H_{R}=\left[
\begin{array}{cccc}
\varepsilon _{+}^{a}(k) & 0 & 0 & -\alpha d \\
0 & \varepsilon _{-}^{a}(k) & \alpha d & 0 \\
0 & \alpha d & \varepsilon _{+}^{b}(k) & 0 \\
-\alpha d & 0 & 0 & \varepsilon _{-}^{b}(k)%
\end{array}%
\right] ,  \label{eq4}
\end{equation}
where $d\equiv \langle \phi _{a}(y)|\partial /\partial y|\phi _{b}(y)\rangle
$, $\varepsilon _{s}^{n}(k)=\hbar ^{2}\left( k-sk_{R}\right) ^{2}/2m^{\ast
}+\epsilon _{n}-\epsilon _{R}$, $\epsilon _{R}\equiv \hbar
^{2}k_{R}^{2}/2m^{\ast }$, ($s=\pm $, $n=a,b$) and we have considered $%
|\sigma \rangle $ to be the eigenbasis of $\sigma _{y}$. For $d=0$ the
Hamiltonian in (\ref{eq4}) is diagonal and yields uncoupled Rashba
dispersions $\varepsilon _{s}^{n}(k)$ (thin lines in Fig. 2); the
corresponding wave functions are $\varphi _{k,n,s}(x,y)$ (here $|\sigma
\rangle \rightarrow |s=\pm \rangle =\left[ |\uparrow \rangle \pm
i|\downarrow \rangle \right] /\sqrt{2}$). Note that for $d=0$ the bands
cross for some values of $k$.\ For instance, for $k>0$ a crossing occurs at $%
k_{c}=(\epsilon _{b}-\epsilon _{a})/2\alpha .$ For non-zero
interband coupling $d\neq 0$ \cite{moroz}, we can diagonalize
$H_R$ exactly (see Mireles and Kirczenow in Ref. \cite{moroz}) to
find the new dispersions (thick lines in Fig. 2).

\emph{Bands near }$k_{c}$. Since we are interested in transport with
injection energies near the crossing, we follow below a simpler perturbative
approach \cite{am} to determine the energy dispersions and wave functions
near $k_{c}$. Near the crossing we can solve the reduced Hamiltonian%
\begin{equation}
\tilde{H}_{R}=\left[
\begin{array}{cc}
\varepsilon _{-}^{a}(k) & \alpha d \\
\alpha d & \varepsilon _{+}^{b}(k)%
\end{array}%
\right] ,  \label{eq5}
\end{equation}%
which to lowest order yields%
\begin{equation}
\varepsilon _{\pm }^{\text{approx}}(k)=\frac{\hbar ^{2}k^{2}}{2m^*}+\frac{1}{2}%
\epsilon _{b}+\frac{1}{2}\epsilon _{a}\pm \alpha d.  \label{eq6}
\end{equation}%
As shown in the inset of Fig. 2, Eq. (\ref{eq6}) describes very well the
anti-crossing of the bands near $k_{c}$.\ The corresponding \emph{zero-order}
eigenstates are
\begin{equation}
|\psi _{\pm }\rangle =\frac{1}{\sqrt{2}}\left[ |-\rangle _{a}\pm |+\rangle
_{b}\right] =\frac{1}{\sqrt{2}}\left[ \left(
\begin{array}{c}
1 \\
-i%
\end{array}%
\right) _{a}\pm \left(
\begin{array}{c}
1 \\
i%
\end{array}%
\right) _{b}\right] ,
\end{equation}%
where the sub-indices indicate the respective channel. The analytical form
in (\ref{eq6}) allows us to determine the wave vectors $k_{c1}$ and $k_{c2}$
in\ Fig. 2 straightforwardly: we assume $k_{c1}=k_{c}-\Delta /2$ and $%
k_{c2}=k_{c}+\Delta /2$ and solve $\varepsilon _{+}^{\text{approx}%
}(k_{c1})=\varepsilon _{-}^{\text{approx}}(k_{c2})$ (assumed $\sim
\varepsilon _{F}$) to find%
\begin{equation}
\Delta =\frac{2m^*\alpha d}{\hbar
^{2}k_{c}}=2\frac{k_{R}}{k_{c}}d\text{.} \label{eq7}
\end{equation}%
Note that to the lowest order used here the horizontal splitting $\Delta $
is constant and symmetric about $k_{c}$.

\emph{Boundary conditions.} We now consider a spin-up electron entering the
Rashba-active region of length $L$ in the wire. Following the usual
approach, we expand this incoming state in terms of the coupled Rashba
states in the wire. We consider only the states $k_{c1}$, $k_{c2}$, and $%
k_{2}$ in the expansion%
\begin{equation}
|\Psi \rangle =\frac{1}{2}|\psi _{+}\rangle e^{ik_{c1}x}+\frac{1}{2}|\psi
_{-}\rangle e^{ik_{c2}x}+\frac{1}{\sqrt{2}}|+\rangle _{a}e^{ik_{2}x}.
\label{eq8}
\end{equation}%
The above \emph{ansatz }satisfies the boundary conditions for both the wave
function and (to leading order) its derivative $x=0$. More explicitly, the
velocity operator condition \cite{molenkamp} at $x=0$ for an electron with $k=k_{F}$ yields%
\begin{equation}
\left(
\begin{array}{c}
k_{F} \\
0 \\
0 \\
0%
\end{array}%
\right) =\frac{1}{2}\left(
\begin{array}{c}
k_{c}+k_{2} \\
-i\left( k_{c}-k_{2}-2k_{R}\right) \\
-\Delta /2 \\
-i\Delta /2%
\end{array}%
\right) =\frac{1}{2}\left(
\begin{array}{c}
k_{c}+k_{2} \\
0 \\
-\Delta /2 \\
-i\Delta /2%
\end{array}%
\right) ,  \label{eq9}
\end{equation}%
where we used $k_{2}-k_{c}=2k_{R}$ (still valid to leading order). The\
`four-vector' notation in (\ref{eq9}) concisely specifies the spin states in
channels \emph{a} (upper half) and \emph{b} (lower half). Note that Eq. (\ref%
{eq9}) is satisfied provided that $\Delta \ll 4k_{F}$.\ This inequality is
satisfied in our system for realistic parameters.

Underlying the \emph{ansatz} in (\ref{eq8}) is the assumption of unity
transmission through the Rashba region. Here we have in mind the particular
spin-transistor geometry sketched in Fig. 1(a): a gate-controlled
Rashba-active region of extension $L$ \emph{smaller} than the total length $%
L_{0}$ of the wire. In this configuration, there are only small band offsets
(which we neglect)\ of the order of $\epsilon _{R}\ll \varepsilon _{F}$ at
the entrance $(x=0)$ and exit $(x=L)$ of the Rashba region. Hence
transmission is indeed very close to unity, see Ref. [\onlinecite{egd}]. The
boundary conditions at $x=L$ are also satisfied.

\emph{Generalized spin-rotated state.}\ From Eq. (\ref{eq8}) we find that a
spin-up electron entering the Rashba region at $x=0$ emerges from it at $x=L$
in the spin-rotated state 
\begin{eqnarray}
\Psi _{\uparrow ,L} &=&\frac{1}{4}\left[ \left(
\begin{array}{c}
e^{-iL\Delta /2} \\
-ie^{-iL\Delta /2} \\
e^{-iL\Delta /2} \\
ie^{-iL\Delta /2}%
\end{array}%
\right) +\left(
\begin{array}{c}
e^{iL\Delta /2} \\
-ie^{iL\Delta /2} \\
-e^{iL\Delta /2} \\
-ie^{iL\Delta /2}%
\end{array}%
\right) \right] e^{ik_{c}L}+\frac{1}{2}\left(
\begin{array}{c}
1 \\
i \\
0 \\
0%
\end{array}%
\right) e^{ik_{2}L}  \notag \\
&=&\frac{1}{2}e^{i(k_{c}+k_{R})L}\left(
\begin{array}{c}
\cos (\theta _{d}/2)e^{-ik_{R}L}+e^{ik_{R}L} \\
-i\cos (\theta _{d}/2)e^{-ik_{R}L}+ie^{ik_{R}L} \\
-i\sin (\theta _{d}/2)e^{-ik_{R}L} \\
\sin (\theta _{d}/2)e^{-ik_{R}L}%
\end{array}%
\right) ,  \label{eq10}
\end{eqnarray}%
%
%
which is essentially Eq. (\ref{eq2}). Observe that in absence of interband
coupling (i.e., $\theta _{d}=0$) Eq. (\ref{eq10}) reduces to the Datta-Das
state in (\ref{eq1}). An expression similar to (\ref{eq10}) holds for the
case of an incoming spin-down electron.

\emph{Spin-resolved current.} For $x\geqslant L$ 
we have%
\begin{eqnarray}
\Psi _{\uparrow }(x &\geqslant &L,y)=\frac{1}{2}\left(
\begin{array}{c}
e^{-i\theta _{R}/2}\cos \left( \theta _{d}/2\right) +e^{i\theta _{R}/2} \\
-ie^{-i\theta _{R}/2}\cos \left( \theta _{d}/2\right) +ie^{i\theta _{R}/2}%
\end{array}%
\right) e^{i\left( k_{c}+k_{R}\right) x}\phi _{a}(y)+  \notag \\
&&\frac{1}{2}\left(
\begin{array}{c}
-ie^{i\theta _{R}/2}\sin \left( \theta _{d}/2\right)  \\
e^{i\theta _{R}/2}\sin \left( \theta _{d}/2\right)
\end{array}%
\right) e^{i\left( k_{c}-k_{R}\right) x}\phi _{b}(y),  \label{eq11}
\end{eqnarray}%
%
which describes planes waves in the \textit{uncoupled} channels \emph{a} and
\emph{b} arising for an incoming spin-up electron in channel \emph{a}. The
total current follows straightforwardly (Landauer-B\"{u}ttiker) from Eq. (%
\ref{eq11})
\begin{equation}
I_{\uparrow ,\downarrow }=\frac{e}{h}eV[1\pm \cos (\theta _{d}/2)\cos \theta
_{R}].  \label{eq12}
\end{equation}%
where $eV\ll \varepsilon _{F}$ is the applied bias between the
source and drain. The spin-dependent conductance in (\ref{eq3})
follows immediately from (\ref{eq12}). Equation (\ref{eq12})
clearly shows the additional modulation $\theta _{d}$ of the
spin-resolved current due to s-o
induced interband coupling. Figure 3 illustrates the angular dependence of $%
G_{\downarrow }$ as a function of $\theta _{R}$ and $\theta _{d}$. The s-o
mixing angle $\theta _{d}$ enhances the possibilities for spin control in
the Datta-Das transistor.

\textit{Realistic parameters.} For concreteness, let us consider infinite
transverse confinement (width $w$). In this case, $\epsilon _{b}-\epsilon
_{a}=3\hbar ^{2}\pi ^{2}/2mw^{2}$ and the interband coupling constant \ $%
d=8/3w$. We choose $\epsilon _{b}-\epsilon _{a}=16\epsilon _{R}$, which
implies (i) $\alpha =(\sqrt{3}\pi /4)\hbar ^{2}/mw=3.45\times 10^{-11}$ eVm
(and $\epsilon _{R}\sim 0.39$ meV) for $m=0.05m_{0}$ and $w=60$ nm, (ii) $%
\varepsilon (k_{c})=24\epsilon _{R}$ [$\varepsilon _{F}$ should be tuned to $%
\sim $ $\varepsilon (k_{c})$], and (iii) $k_{c}=8\epsilon _{R}/\alpha $.
Assuming $L=69$ nm [Rashba region length, Fig. 1(a)], we find $\theta
_{R}=\pi $ and $\theta _{d}=\theta _{R}d/k_{c}=\pi /2$, since $d/k_{c}\sim
0.5$. This is a conservative estimate. In principle, $\theta _{d}$ can be
varied independently of $\theta _{R}$ via lateral gates which alter $w$.
Note also that $\Delta /4k_{F}\sim 0.05$ [validity of Eq. (\ref%
{eq9})] for the above parameters. Finally, we note that the most
relevant spin-flip mechanism (Dyakonov-Perel) should be suppressed
in quasi-one-dimensional systems such as ours \cite{bournel}. In
addition, thermal effects are irrelevant in the experimentally
feasible linear regime \cite{datta-book} we consider here

This work was supported by NCCR\ Nanoscience, the Swiss NSF, DARPA, and ARO.
We acknowledge useful discussions with D. Saraga.

\newpage

\noindent \textbf{List of references}

\newpage

\noindent \textbf{Figures}

\begin{figure}[th]
\begin{center}
\epsfig{file=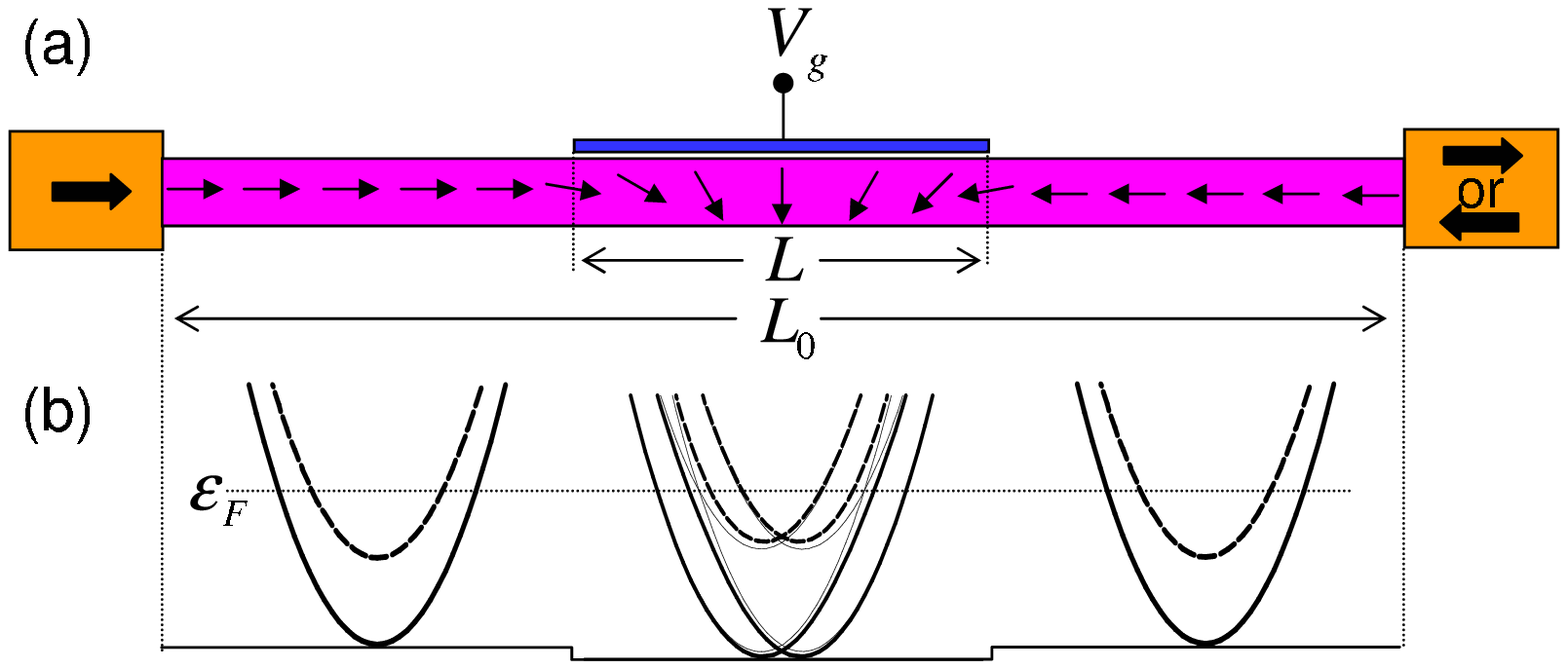, width=0.45\textwidth}
\end{center}
\caption{Spin transistor geometry with a two-band channel. (a) The length $L$
of the Rashba region is smaller than the total length $L_0$ of the wire. (b)
Sketch of energy dispersions in the s-o active region with and without
interband coupling (Rashba bands) and away from it (parabolic bands). Note
the small band offsets between adjacent regions in the wire.}
\label{fig:fig1}
\end{figure}

\begin{figure}[th]
\begin{center}
\epsfig{file=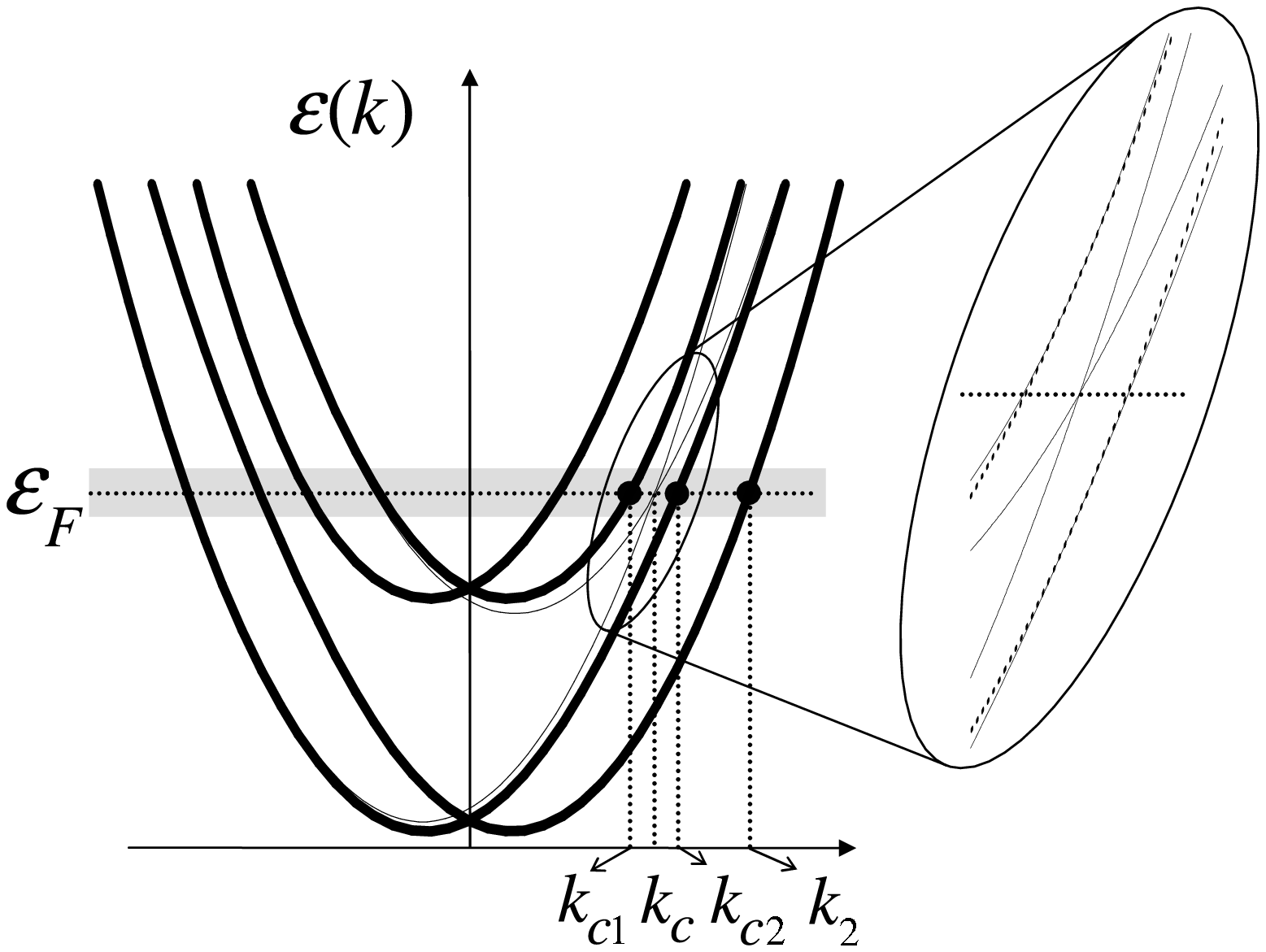, width=0.45\textwidth}
\end{center}
\caption{Band structure in the presence of spin-orbit coupling. In absence
of interband mixing the Rashba dispersions are uncoupled (thin solid lines)
and cross at, e.g., $k_{c}$. For non-zero interband coupling the bands anti
cross (thick solid lines). The inset shows a blowup of the dispersion region
near the crossing: the approximate solution [dotted lines, perturbative
approach, Eq. (\protect\ref{eq6})] describes well the energy dispersions
near $k_{c}$. }
\label{fig:fig2}
\end{figure}

\begin{figure}[th]
\begin{center}
\epsfig{file=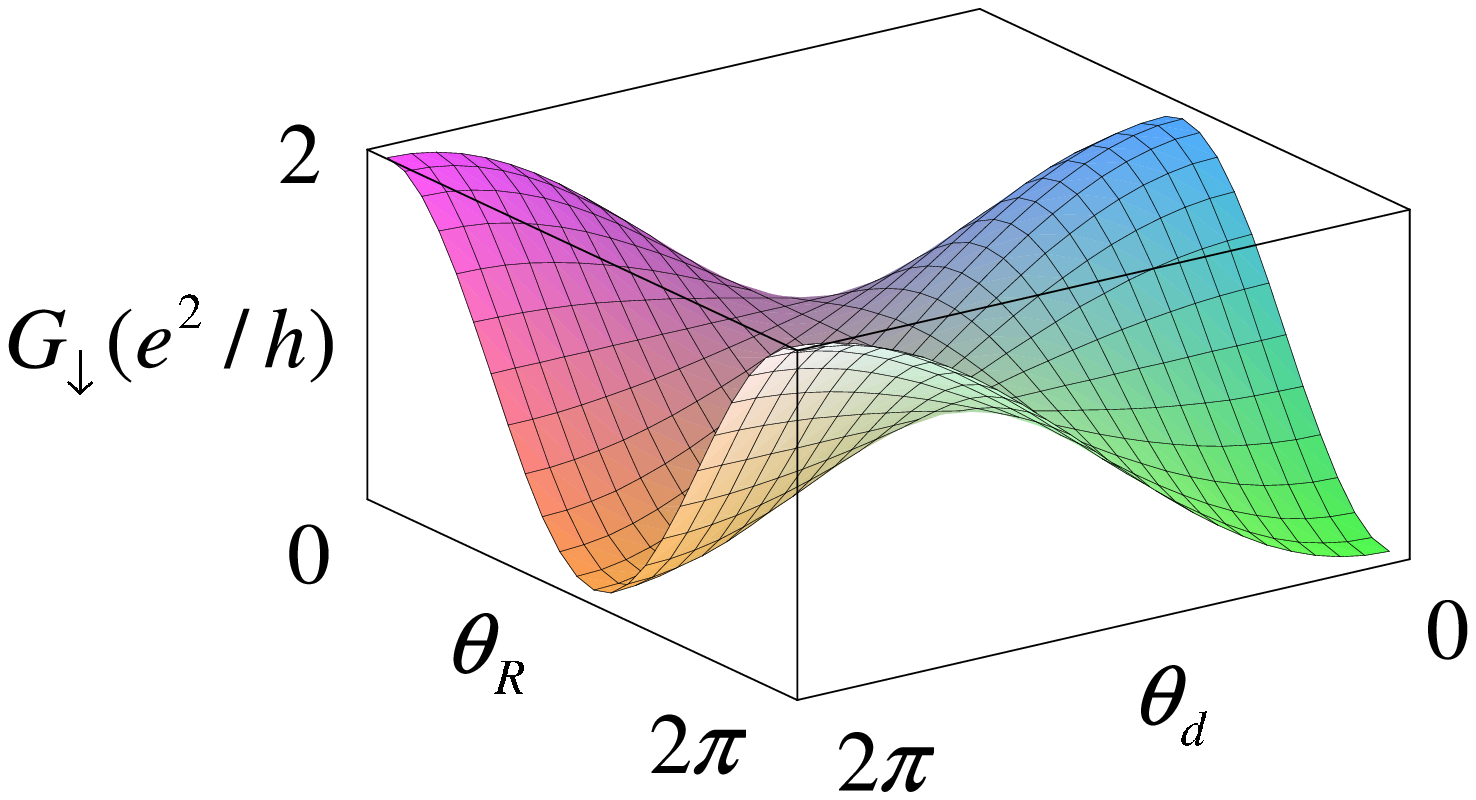, width=0.37\textwidth}
\end{center}
\caption{Angular dependence of the spin-down conductance. The additional
modulation $\protect\theta _{d}$ due to s-o interband mixing and $\protect%
\theta _{R}$ can be varied independently.}
\label{fig:fig3}
\end{figure}


\begin{thebibliography}{99}
\bibitem{datta} S. Datta and B. Das, Appl.\ Phys.\ Lett. \textbf{56}, 665
(1990).

\bibitem{exp} G. Meir, T. Matsuyama, and U. Merkt, Phys. Rev. B \textbf{65}, 125327
(2002); C.-M. Hu, J. Nitta, A. Jensen, J. B. Hansen, H.
Takayanagi, T. Matsuyama, D. Heitmann, and U. Merkt, J. Appl.\
Phys. \textbf{91}, 7251 (2002).

\bibitem{spin-pol} R.\ Fiederling, M. Keim, G. Reuscher, W. Ossau,
G. Schmidt, A. Waag, L. W. Molenkamp, Nature \textbf{402}, 787
(1999); Y. Ohno, D. K. Young, B. Beschoten, F. Matsukura, H. Ohno,
D. D. Awschalom, Nature \textbf{402}, 790 (1999).

\bibitem{egues} J. C. Egues, Phys.\ Rev.\ Lett.\ \textbf{80}, 4578 (1998).

\bibitem{rashba} Yu.\ A. Bychkov and E. I. Rashba, JETP Lett.\ \textbf{39},
78 (1984).

\bibitem{nitta} J. Nitta T. Akazaki, H. Takayanagi, and T. Enoki,
Phys.\ Rev.\ Lett.\ \textbf{78}, 1335 (1997); G. Engels, J. Lange,
Th. Sch\"apers, and H. L\"uth, Phys.\ Rev.\ B \textbf{55}, R1958
(1997); D. Grundler, Phys.\ Rev.\ Lett.\ \textbf{84}, 6074 (2000);
Y. sato, T. Kita, S. Gozu, and S. Yamada, J. Appl. Phys.
\textbf{89}, 8017 (2001).

\bibitem{am} N. W. Ashcroft and N. D. Mermin, \emph{Solid State Physics},
Ch. 9. (Holt, Rinehart, and Winston, New York, 1976).

\bibitem{moroz} A. V. Moroz and C. H.\ W. Barnes, Phys.\ Rev. B \textbf{60},
14272 (1999); F. Mireles and G. Kirczenow, Phys. Rev. B
\textbf{64}, 024426 (2001); M. Governale and U. Z\"{u}licke, Phys.
Rev. B \textbf{66}, 073311 (2002).

\bibitem{egd} J. C. Egues, G. Burkard, and D. Loss,
Phys. Rev. Lett. \textbf{89}, 176401 (2002)).

\bibitem{molenkamp} E. A. de Andrada e Silva, G. C. La Rocca, and F. Bassani, Phys. Rev. B
\textbf{55}, 16293 (1997); L. W. Molenkamp, G. Schmidt, and G.E.W.
Bauer, Phys. Rev. B \textbf{64}, R121202 (2001).

\bibitem{bournel} A. Bournel, P. Dollfus, P. Bruno, and P. Hesto, Eur. Phys. J. AP \textbf{4},
1 (1998).

\bibitem{datta-book} S. Datta, \textit{Electronic transport in mesoscopic
systems} (Cambridge University Press, Cambridge, 1997), Ch. 2, p.
89.
\end{thebibliography}
\end{document}